\documentclass[twocolumn,showpacs,preprintnumbers,amsmath,amssymb,prl]{revtex4-1}

\usepackage[utf8]{inputenc}
\usepackage{graphicx}  
\usepackage{psfrag}    
\usepackage{amsthm}    
\usepackage{makeidx}   
\usepackage{braket}    
\usepackage{pstricks}
\usepackage{tikz}
\usepackage{wrapfig}
\usepackage[english]{babel}
\addto\captionsenglish{}
\usepackage{hyperref}
\usepackage{epstopdf}

\hypersetup{
  colorlinks   = true, 
  urlcolor     = blue, 
  linkcolor    = blue, 
  citecolor   = red, 
}

\title{A Monte Carlo Time-Dependent Variational Principle}

\def\Cx{{\mathbb C}}     
\def\Tw{{\mathbb T}}     

\def\bra #1{\langle #1\vert}
\def\ket #1{\vert #1\rangle}

\def\ketbra #1#2{\vert #1\rangle \langle #2\vert}

\def\tr{\mathop{\rm tr}\nolimits}

\mathchardef\ree="023C \mathchardef\imm="023D  

\def\LL{{\mathcal L}}

\newcommand\numberthis{\addtocounter{equation}{1}\tag{\theequation}}

\makeindex

\begin{document}


%
%
%
%
%
%


\title{A Monte Carlo Time-Dependent Variational Principle}

\author{F.\,W.\,G. Transchel} 
 \email{fabian.transchel@itp.uni-hannover.de}
 \affiliation{Institut f\"ur Theoretische Physik, Appelstr. 2, Hannover, D-30167, Germany}
 
\author{A.\,Milsted}%
 \email{ashley.milsted@itp.uni-hannover.de}
 \affiliation{Institut f\"ur Theoretische Physik, Appelstr. 2, Hannover, D-30167, Germany}

\author{Tobias\,J.\,Osborne}
 \email{tobias.osborne@itp.uni-hannover.de}
 \affiliation{Institut f\"ur Theoretische Physik, Appelstr. 2, Hannover, D-30167, Germany}

\date{\today}

\begin{abstract}
We generalize the Time-Dependent Variational Principle (TDVP) to dissipative
systems using Monte Carlo methods, allowing the application of existing variational
classes for pure states, such as Matrix Product States (MPS), to the simulation
of Lindblad master equation dynamics. The key step is to use sampling to approximately
solve the Fokker-Planck equation derived from the Lindblad generators.
An important computational advantage of this method, compared to other variational approaches
to mixed state dynamics, is that it is ``embarrassingly parallel''.
\end{abstract}

\maketitle

\vspace{0.5cm}


Quantum many body systems are hard to solve. 
Even the largest supercomputers are stumped by the general case because
the dimension of Hilbert space scales exponentially with the number of particles.
However, most of this vast space corresponds to states that are highly \emph{entangled}, 
a property which is \emph{not} possessed by a great many 
physically relevant states \cite{Eisert2010, hastings_area_2007, osborne_efficient_2006}.
Fortunately, this can be exploited by working with \emph{variational classes} of states that 
parameterize this highly relevant corner of Hilbert space.
In the case of one-dimensional systems, this is achieved by the hugely successful 
Density Matrix Renormalization Group (DMRG) technique
\cite{White1992, White1993, peschel1999}, which 
can be understood \cite{schollwock_density-matrix_2011} as a variational method 
based on Matrix Product States (MPS) | a class for which entanglement is 
upper-bounded by the dimension of the MPS parameter space \cite{Fannes1992, *cirac_renormalization_2009}.
Recently, the Time-Dependent Variational Principle (TDVP) has also been applied 
to MPS, providing a very promising framework for finding ground states, the 
simulation of dynamics, and for probing the excitation spectrum
\cite{haegeman_tdvp_mps, haegeman_variational_2012, Milsted2012}.

Although these advances have revolutionized the simulation of many body systems, 
most work so far has focused on unitary dynamics. Deviations
from unitarity are generally from undesired couplings to the environment that
destroy coherence and are, to that extent, merely an experimental nuisance.
However, it has become increasingly clear that introducing dissipation 
in a controlled way by engineering the system-environment coupling
can be instrumental in performing a number of very 
useful tasks in quantum information processing
\cite{barreiro_open-system_2011, *krauter_entanglement_2011, 
diehl_quantum_2008, *kraus_preparation_2008, *verstraete_quantum_2009,
*eisert_noise-driven_2010,*kastoryano_dissipative_2011, 
*muschik_dissipatively_2011, *pastawski_quantum_2011, 
*vollbrecht_entanglement_2011, *mari_cooling_2012}. 
From this perspective, being
able to efficiently simulate dissipative dynamics is highly desirable.

Existing methods for the simulation of general dissipative dynamics are based 
either on Monte Carlo sampling or on variational classes of mixed states. 
On one hand, sampling wavefunctions \cite{gardiner_noise, gisin1992} is an
easily parallelizable problem that is, however, in general limited to small systems
due to the computational demands of evolving wavefunctions for each sample trajectory.
On the other, general variational methods such as the TDVP for mixed states 
\cite{kraus_tdvp_diss_2012} suffer from the lack of a unique measure 
of information distance for comparing mixed states. More specific techniques, 
such as those using MPS on purified states with time-evolving block decimation 
\cite{Verstraete2004, zwolak2004} are somewhat less easy to parallelize. 
A natural way of combining these two approaches is to choose a variational class 
of pure states to represent the wavefunction components of a density matrix, 
using Monte Carlo sampling to simulate dissipation. This was suggested in \cite{Verstraete2004}
and applied with a mean field wavefunction ansatz in \cite{pichler_heating_2013}, 
yet the use of more complicated variational classes in this way
has so far focused on the specific case of generating approximations 
of thermal states \cite{white_METTS_2009, *Stoudenmire2010, garnerone2013}.

In this Letter, we introduce a general variational approach to dissipative 
dynamics that makes use of an arbitrary pure state variational class, thus 
avoiding the need to explicitly fix a measure of information distance for
mixed states. We do this by applying Monte Carlo sampling to the TDVP flow
equations derived from the Lindblad master equation describing the system.
The use of sampling creates an effective variational class of mixed states
from the pure state ansatz and has the added benefit of making the resulting 
algorithm ``embarrassingly parallel''\cite{breen_invitation_2008}.
We implement the method for MPS and trial it on a simple spin-chain system
with nearest-neighbor interactions and spin-flip dissipation to check 
convergence of the sampling. We then test it 
on a larger XXZ Heisenberg chain driven at the edges, a system with known analytic
solutions.





\begin{figure}[htbp]
      \begin{minipage}[t]{0.45\textwidth}
        \includegraphics[width=7.5cm]{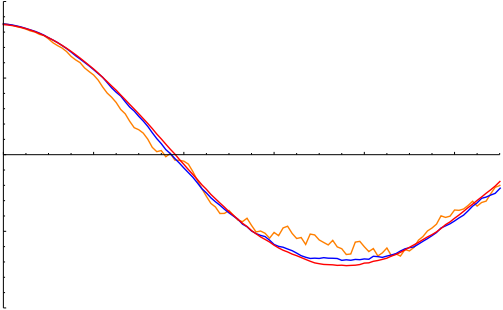} 
	\vspace{.5cm}
	\caption{Comparison of the accuracy of an ensemble of MPS representations at 
	  480 (yellow), 4800 (blue), 48000 (red) samples. The exact solution
	  is not shown because the difference from the red curve would not be
	  visible on this plot.
	  We plot from $t=80$ because the difference between the curves is too small
	  to be seen for $t < 80$. 
	}
	\label{fig1}
       \end{minipage}
\end{figure}







We begin in a general dissipative setting where the dynamics of a state $\rho$, 
belonging to a Hilbert space $\mathcal{H}$,
is given by a Lindblad master equation of the form 
\begin{equation}
 \partial_t \rho = - \mathrm{i} [K,\rho] - \frac{1}{2}\sum_\alpha({L_\alpha}^\dagger L_\alpha \rho + \rho {L_\alpha}^\dagger L_\alpha - 2 L_\alpha \rho L_\alpha^\dagger),
 \label{eq1}
\end{equation} 
where $K$ is Hermitian, $L_\alpha$ are bounded operators 
and we can write the RHS as $Q\rho + \rho Q^\dagger + \sum_\alpha L_\alpha \rho L_\alpha^\dagger$
using $Q = -\mathrm{i}K - \frac{1}{2}\sum_\alpha {L_\alpha}^\dagger L_\alpha$.
We approximate a general mixed state at a time $t$ as
\begin{equation}
  \rho_t = \int_{\mathcal{M}} p_t(\bar{a}, a) \ketbra{\Psi(a)}{\Psi(a)} da d\bar{a},
\label{eq2}
\end{equation}
where $\ket{\Psi(a)}$ belongs to a variational class of wavefunctions with parameters $a$, such as MPS for one-dimensional
lattice systems, and $p_t(\bar{a}, a)$ is a probability distribution over those parameters.

If we come back to the integral from before, note that it is over the submanifold of Hilbert space $\mathcal{M} \in \mathcal{H}$ formed by
$\ket{\Psi(a)} \in \mathcal{M}$ and cannot generally be evaluated efficiently since the dimension
of $\mathcal{M}$ is, despite being much smaller than $\dim (\mathcal{H})$, still large. 
However, as we show, this ansatz lends itself naturally to Monte Carlo sampling.
In the following, we assume for reasons of simplicity and 
without loss of generality, that $\ket{\Psi(a)}$ depends holomorphically on $a$.

Applying the master equation \eqref{eq1} to the ansatz \eqref{eq2} gives us an infinitesimal time step 
$\partial_t\rho_t$ in terms of the evolution $\partial_t p_t$ of the probability distribution
together with the action of the operators $Q$ and $L_\alpha$ on the pure state $\ket{\Psi(a)}$.
If we wish to maintain the form \eqref{eq2} whilst evolving the state, 
the latter must be approximated by a vector $\ket{\Phi(b)} \equiv b^j \partial_{a^j} \ket{\Psi(a)}$ 
in the tangent space $\Tw_a$ to the variational manifold $\mathcal{M}$ at point $a$
(note that repeated indices are summed over unless otherwise stated).
This kind of approximation forms the basis of the time-dependent variational principle (TDVP), 
as explained further in the supplementary material. 
The tangent vector parameters $b_Q$ optimally approximating $Q\ket{\Psi(a)}$ can be found 
via $b_Q \approx \arg \min_{b'} \left| Q\ket{\Psi(a)} - \ket{\Phi(b')} \right|^2$, with
$b_\alpha$ for each $L_\alpha$ following in identical fashion.
After partial integration (discarding surface terms), we obtain an effective 
master equation
\begin{align*}
\partial_t \rho_t = \! \int_{\mathcal{M}} \!\left[-\partial_{a^j} (p_t b_Q^j) - \text{c.c.} + \partial_{a^k} \partial_{\bar{a}^l}(p_t b^k_\alpha \bar{b}^l_\alpha)\right]\!
\\ \ketbra{\Psi(a)}{\Psi(a)} da d\bar{a}, \numberthis
\end{align*}
which evolves states only within the mixed state ansatz \eqref{eq2},
approximating the exact Lindblad dynamics in a locally optimal (in time) way.
We now use sampling to evaluate the integral by taking the variational parameters
$a$ to be random variables, which may lead
to a solution (see \cite{Gardiner2002})
in form of a stochastic differential equation (SDE) 
\begin{equation}
  \label{eqn:tdvp_sde}
  da^j(t) = \left(b_{Q}^j + \langle\bar{L}_\alpha\rangle b_\alpha^j\right) dt + b^j_\alpha dw_\alpha(t).
\end{equation}
These are identical to the TDVP flow equations \cite{haegeman_tdvp_mps}
with additional dissipative noise captured by the $b_\alpha$
and can be sampled from via numerical integration beginning from some starting parameters $a_0$.
There exist other ways to unravel a master equation, including quantum jumps (see \cite{Daley2014}, \cite{Bonnes2014}).
However the Quantum State Diffusion method \cite{gisin1992} is best suited to the TDVP.
If the variational manifold $\mathcal{M}$ captures the full Hilbert space, the presented method reduces to the exact QSD.
Expectation values for $\rho_t$ can be computed approximately with the standard convergence of 
$1 / \sqrt{N}$, where $N$ is the number of samples.
\begin{figure*}[htbp] 
    \includegraphics[width=8cm]{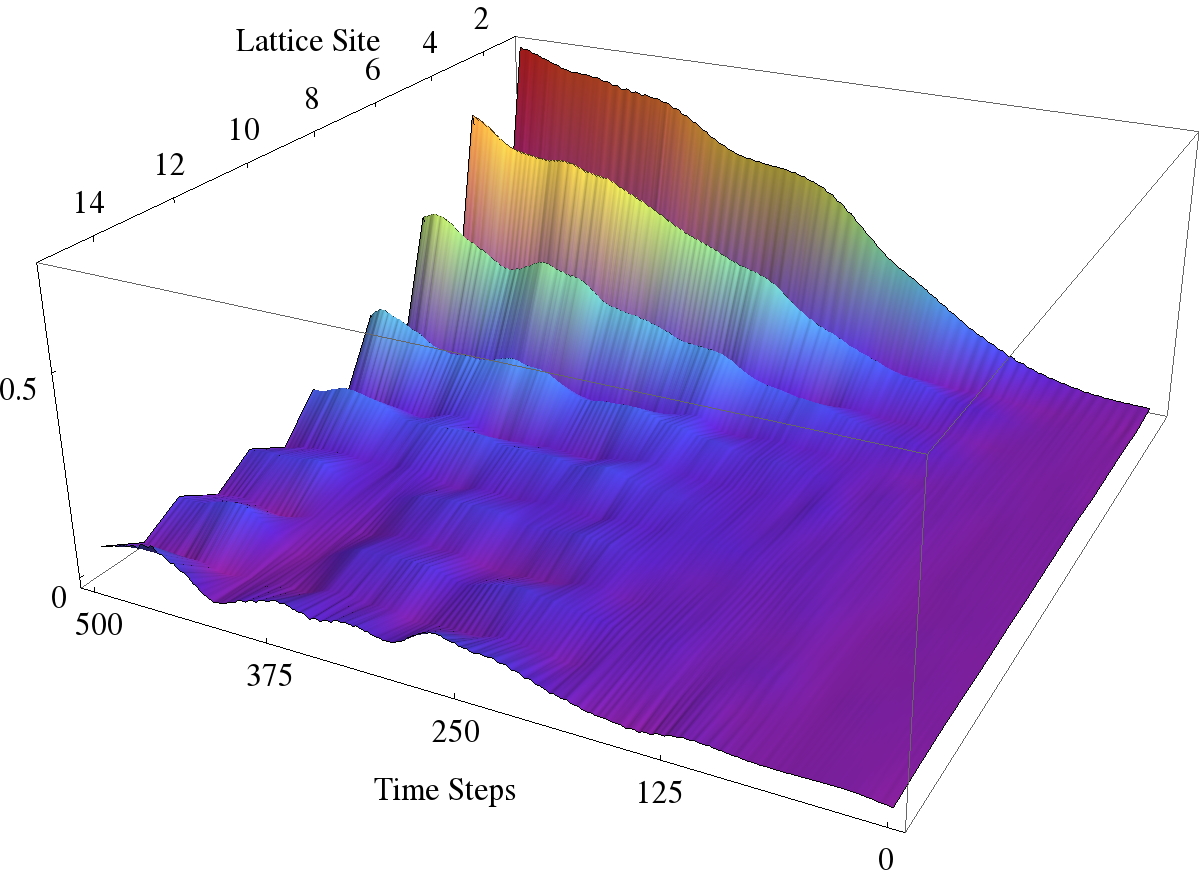}
    \hspace{.5cm}
    \includegraphics[width=8cm]{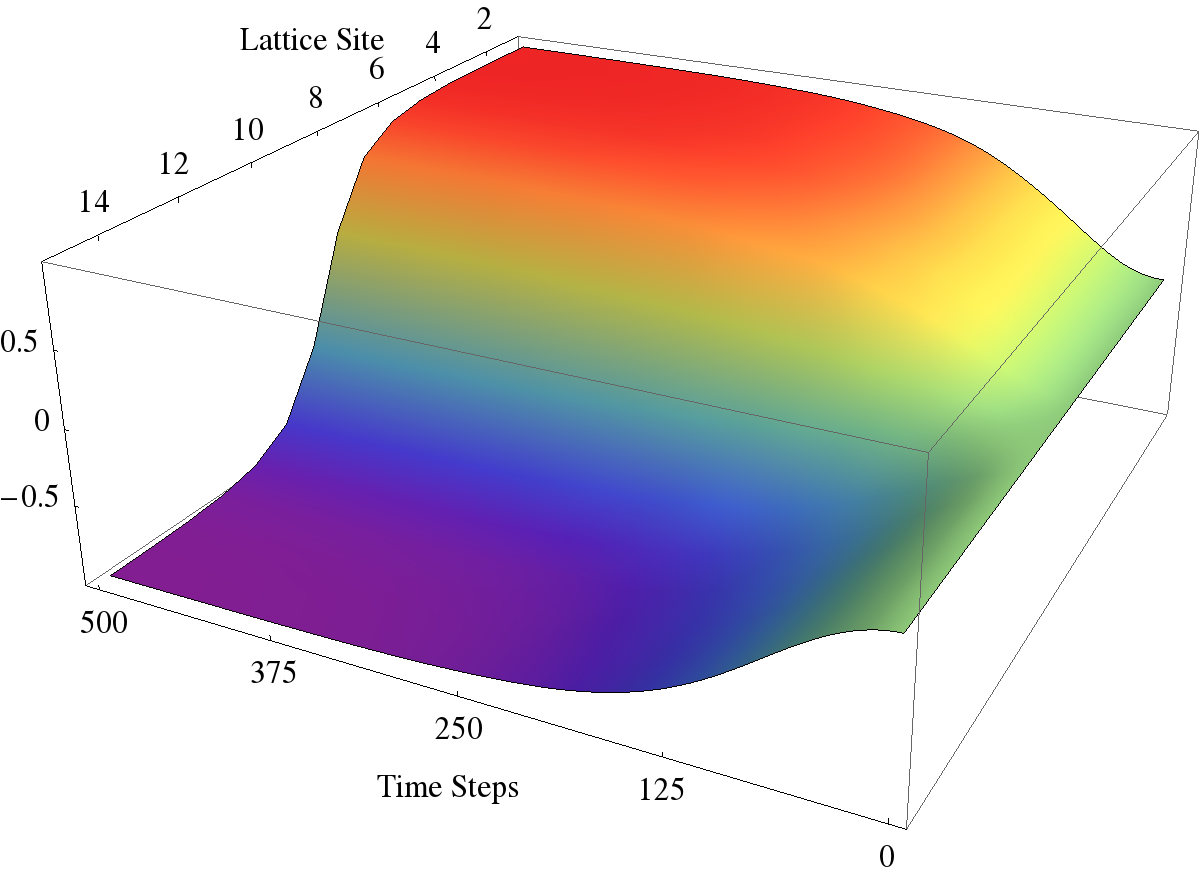}
    \caption{\textbf{Left} Time-resolved two-point $\langle\sigma^x \sigma^x\rangle$-correlation function (vertical axis) of site 8 with all other sites
    for the Heisenberg-type model $K_{XZ}$ (see text) under spin-flip dissipation $L = \sum_n \sigma^+_n$,
    simulated using $n = 16$ sites, $N = 1000$ samples and bond dimensions $D=32$.
    We clearly observe that the spin correlations comply with the expected antiferromagnetic domain behaviour of the $K_{XZ}$ model,
    while the absolute values of the spin expectations are skewed due to the inherent symmetry breaking induced by the finite length of the lattice.
    \textbf{Right} Same plot for bihomogenous dissipation of the form
    $L_{\alpha = {1\dots \frac{n}{2}}}=\sigma^+_\alpha, L_{\alpha = {(\frac{n}{2} + 1) \dots n}} = \sigma^-_\alpha$
    on a Heisenberg XZ lattice with $n = 16$, $N = 1000$ and $D=64$. Correlations are smoothed throughout the lattice due to the fact that
    the spin alignment is mediated between the two antipodal domains. Note that this was not at all clear at the domain
    flip region in the center of the lattice.
    }
    \label{fig2}
\end{figure*}

Numerical integration can be performed for each sample using, for example, the following
algorithm implementing the Euler method:
\begin{enumerate}
 \item Generate a starting state for the sample by initializing the variational parameters $a^j$ with suitable values.
 \item To evolve from time $t$ to $t+dt$, evaluate \eqref{eqn:tdvp_sde} and set $a^j(t+dt) = a^j(t) + da^j(t)$.
 \item Normalize $\ket{\Psi(a)}$ if necessary and restore a canonical form for $a^j$ as needed.
 \item Calculate expectation values of interest, i.e.\,energy or magnetization of the lattice or its elements and go to step 2.
 How one would calculate such ensemble expectations is also explained in the supplementary material.
\end{enumerate}
The need for normalization and the restoration of a canonical form varies depending on the chosen
pure state variational class. For example, in the case of MPS the first is needed and, due to 
redundancy in the choice of parameters, the second is recommended for numerical conditioning.

Since samples are completely independent of one another they can be evaluated simultaneously using any
sufficiently capable computer processors, making this method ``embarrassingly parallel'' with regard
to scaling in $N$. This is fortunate, since the variance of approximate expectation values scales with 
$\frac{1}{\sqrt{N}}$, such that we would need to square the number of sample runs to double the accuracy. 
For our tests, we thought about sending phishing mails to fellow physicists to take over their personal computers as
computational zombie nodes, but were ultimately content with the existing computation servers available for use at our institute.



%
%

We implemented the method using matrix product states (MPS) as the 
pure state variational ansatz, which is a class well-suited to approximating states 
of many-body systems in one dimension,
where the Hilbert space is taken to be isomorphic to 
$\mathcal{H} = (\mathbb{C}^d)^{\otimes N_\text{sites}}$, 
with a local dimension of $d$.
MPS for finite $N_\text{sites}$ with open boundary conditions have the form 
$\ket{\Psi[A(t)]} = \sum_{s=1}^d {v_L}^\dagger A_1^{s_1}\ldots A_N^{s_N} v_r \ket{s}$, 
where $\ket{s} = \ket{s_1\ldots s_N}$, $A_n^s \in M_D(\Cx)$ and ${v_L}^\dagger$, $v_R$
are boundary vectors that may be absorbed into $A_1$ and $A_N$ respectively.
The corresponding variational manifold $\mathcal{M}_\text{MPS(D)}$ 
constitutes a 
submanifold of $\mathcal{H}$, where $D$ is the so-called bond-dimension |
the maximum Schmidt rank of the Schmidt-decompositions made by cutting
between any two neighboring sites. 
We implement the method by extending evoMPS \cite{evoMPS}, 
an open-source implementation of the TDVP for MPS,
calculating the MPS tangent vectors approximating
local $Q$ (consisting of, say, nearest-neighbor terms) 
in the same way as for a local Hamiltonian \cite{haegeman_tdvp_mps}. 
It is sufficient for the following examples to restrict $L_\alpha$ 
to on-site operators. Since applying an on-site operator to an MPS
results in a tangent vector ($L_\alpha\ket{\Psi[A]} \in \mathbb{T}_{[A]}$), 
no approximation is needed in calculating the $b_\alpha$.


\begin{figure}[h]
	  \includegraphics[width=.45\textwidth]{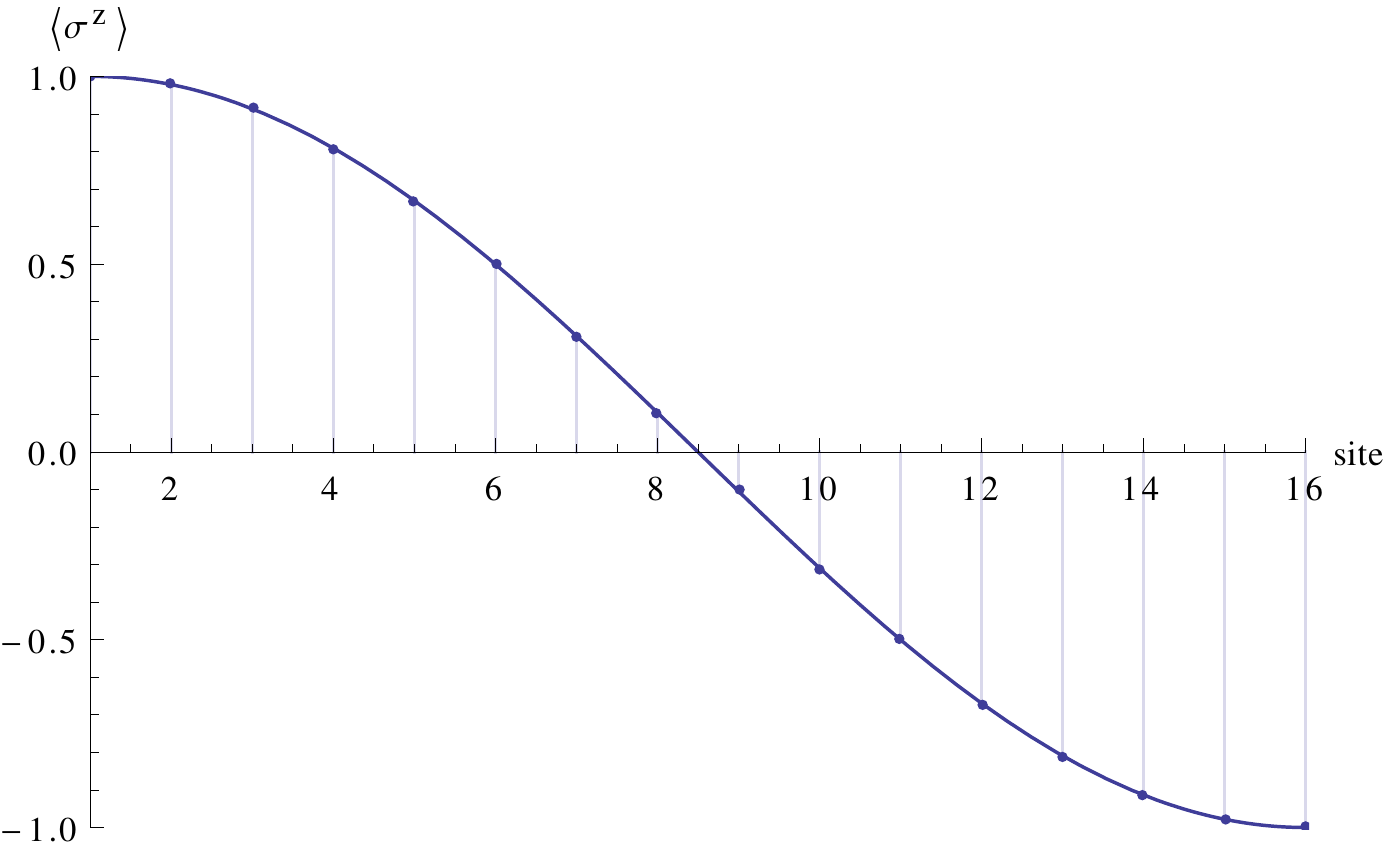}
 \caption{Numerical results for the XXZ chain with edge driving ($n_\text{sites}=16, \epsilon = 1, \lambda = 1,$ $D = 24$), 
          showing good agreement with the analytical solution in \cite{prosen2011} even for a very low number of samples (300).}
  \label{fig:XXZ}
\end{figure}

\begin{figure}[h]
	  \includegraphics[width=.45\textwidth]{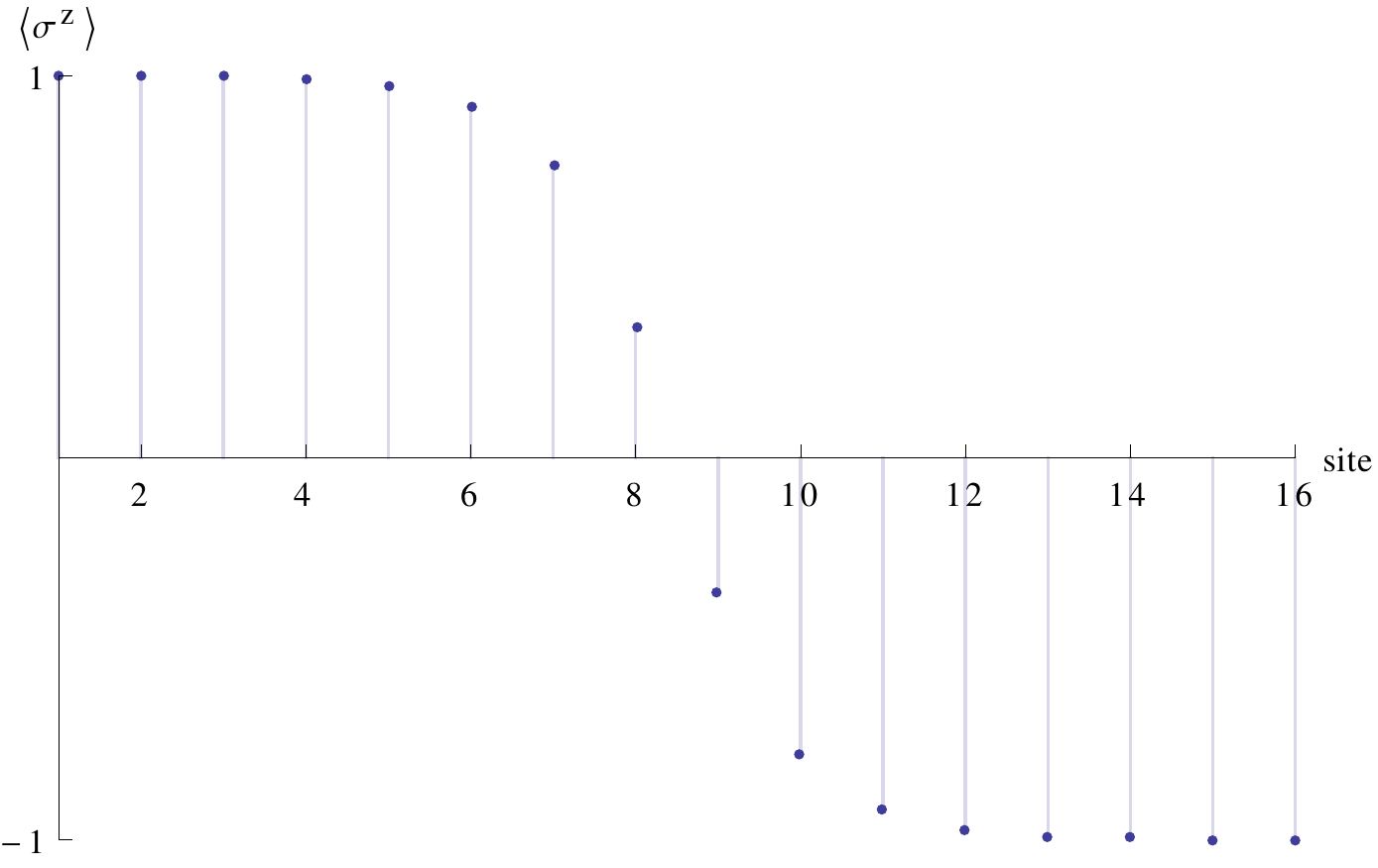}
 \caption{Bihomogenous dissipation for the XXZ chain at low interaction strength
 (or strong driving, $\epsilon = 10^{-3}$, $\lambda = 1$, $n = 16$, $N = 2500$, $D = 64$)}
  \label{fig:XXZbihom}
\end{figure}


To test the method we compare it to an exact calculation of dissipative dynamics 
on a two-qubit ($N_\text{sites} = 2$) lattice $\mathcal{H} = \mathbb{C}^2 \otimes \mathbb{C}^2$.
We define single site Lindblad operators $L_\alpha = \sigma_\alpha^+$ for $\alpha = 1 \dots N_\text{sites}$
and a Hamiltonian part
$K_{XZ} = \sum_{n=1}^{N_\text{sites} - 1} \sigma_n^x \sigma_{n+1}^x + \lambda \sigma_n^z \sigma^z_{n+1}$
and simulate the dissipative dynamics, beginning with a highly entangled MPS
(within the constraints of the chosen bond dimension).
We can readily check that a full state tomography of the two-site density matrix 
differs from the exact analytic solution only by the expected variance of 
$\frac{1}{\sqrt{N}}$, where $N$ is the number of sample paths calculated. 
Figure \ref{fig1} shows convergence with increasing $N$.
We are also able to replicate Rabi oscillations between the spin components of 
the two sites in the $x$ direction.
We conclude that our method and implementation are technically and numerically 
robust enough to be trialed on larger systems, beyond the reach of exact
numerics.


We simulated the same dynamics as above on a longer spin chain 
with $N_\text{sites}=16$, finding that the clustering of correlations grows
as we increase the bond dimension $D$ to match the number of degrees
of freedom the system needs to to approximate the dynamics to good accuracy.
Figure \ref{fig2} shows the time evolution of the
$\sigma^x$ two-point correlation functions (using the 8th site as the reference)
for bond dimensions $D=2$ and $D=32$.
Correlation flow is somewhat comparable to Rabi oscillations in a 2-qubit system.


An XXZ Heisenberg model of the form $K = \sum_{n=1}^{N_\text{sites} - 1} 
\epsilon[ 2\sigma^+_n\sigma^-_{n+1} + 2\sigma^-_n\sigma^+_{n+1} + \lambda\sigma^z_n\sigma^z_{n+1} ]$
(with $\epsilon$ controlling the relative strength of the site coupling 
with respect to the dissipation) 
has been analytically solved for strong driving
in \cite{prosen2011}. It is thus suitable for demonstrating the method
on larger system sizes where the use of a pure state variational ansatz 
like MPS leads to large advantages over other Monte Carlo methods.
When dissipation is introduced by Lindblad operators 
$L_1 = \sigma_1^+, L_2 = \sigma_N^-$ acting on the ends of the lattice, we find
(see figure \ref{fig:XXZ})
that our method can replicate analytic results at small lattice sizes at the 
expense of high sample numbers.
For larger $N$ however, we clearly observe errors made due to the chosen bond 
dimension becoming a limiting factor.
In these cases the information about dissipation at the far ends of the system 
does not permeate to the center, where we observe the largest deviation from 
analytic results | the center spins don't appear to be coupled to the environment at all. 




It should be noted that convergence is expected to be slow for this model because the dynamics are critical \cite{xxz_criticality}.
The information about the center of the chain propagates only slowly to the reservoir at the edges
and thus renders the XXZ chain with edge pumping in some sense the worst kind of system one could imagine for this method.
With that in mind it seems remarkable that the results
are at least qualitatively comparable to the analytics. 


We then explored non-integrable systems like an XXZ chain with bihomogenous dissipation
consisting of a Lindblad operator for each site $L_n = \sigma_n^+$ for $n \le N/2$ 
and $L_m = \sigma_m^-$ for $m > N/2$. 
For large interaction strength, the results are intuitive as there are 
two clearly separate domains
of magnetization in the system, while for weak interactions (see figure \ref{fig:XXZbihom}), especially 
near the center, we can see interference between the up and down pumping of magnetization.
Furthermore we conclude that the bond dimension of the MPS need not approach 
the $2^{N/2}$ needed to represent an arbitrary state exactly in order to capture 
interference effects and maintain sufficient amounts of entanglement throughout the lattice.
This system behaves like a one-dimensional bar magnet in the sense that we are 
able to tune the system parameters $\epsilon$ and $\lambda$ in such a way that we can
explore the behavior at the center, i.e. the zone of spin domain change.
For large magnetic interaction and weak dissipative coupling we find that the method is able to highlight interference patterns near the center.



In this paper we presented a Monte Carlo extension to the TDVP, expanding the scope of its applicability.
In the future it seems natural to apply the method to larger dissipative systems. That includes, but is not
limited to, larger lattice sizes as well as larger internal Hilbert space sizes. Since the method still is, in essence,
a variational approach, it can outperform exact diagonalization while errors stay limited as
long as we can afford to produce enough samples. With the inherent statistical nature of the method, it should be
possible to approximate solutions to great accuracy where analytic solutions are highly non-tractable, 
as long as parallel computing resources are available.
It is also worth noting that the inherent scalability of the method would allow to efficiently use it on even larger
grid scales such as supercomputers to gain insights into previously unreachable domains.
The software developed for this work is based on the evoMPS project \cite{evoMPS} and is freely available 
under an open source license.



\begin{acknowledgments}
This work was supported by the
ERC grants QFTCMPS, and SIQS by the cluster of excellence EXC 201 Quantum Engineering and Space-Time
Research.
F.\,W.\,G. T. wishes to acknowledge useful discussions with R.\,F. Werner, K. Abdelkhalek and B. Neukirchen.
\end{acknowledgments}

\bibliography{ref}




\appendix

\section{Supplementary material\label{supplementary}}

\subsection{Time Dependent Variational Principle}

Given a starting state $\ket{\Psi(a)}$ belonging to a pure state
variational class with parameters $a$ and a differential equation describing 
some dynamics, we wish to compute the approximate time evolution of the state
with the restriction that it must remain in the variational class. 

\begin{figure}[h]

\ifx\du\undefined
  \newlength{\du}
\fi
\setlength{\du}{5\unitlength}
\begin{tikzpicture}
\pgftransformxscale{1.000000}
\pgftransformyscale{-1.000000}
\definecolor{dialinecolor}{rgb}{0.000000, 0.000000, 0.000000}
\pgfsetstrokecolor{dialinecolor}
\definecolor{dialinecolor}{rgb}{1.000000, 1.000000, 1.000000}
\pgfsetfillcolor{dialinecolor}
\pgfsetlinewidth{0.100000\du}
\pgfsetdash{}{0pt}
\pgfsetdash{}{0pt}
\pgfsetbuttcap
{
\definecolor{dialinecolor}{rgb}{0.000000, 0.000000, 0.000000}
\pgfsetfillcolor{dialinecolor}
\pgfsetarrowsend{stealth}
\definecolor{dialinecolor}{rgb}{0.000000, 0.000000, 0.000000}
\pgfsetstrokecolor{dialinecolor}
\draw (8.000000\du,21.000000\du)--(8.000000\du,3.000000\du);
}
\pgfsetlinewidth{0.100000\du}
\pgfsetdash{}{0pt}
\pgfsetdash{}{0pt}
\pgfsetbuttcap
{
\definecolor{dialinecolor}{rgb}{0.000000, 0.000000, 0.000000}
\pgfsetfillcolor{dialinecolor}
\pgfsetarrowsend{stealth}
\definecolor{dialinecolor}{rgb}{0.000000, 0.000000, 0.000000}
\pgfsetstrokecolor{dialinecolor}
\draw (6.000000\du,19.000000\du)--(29.000000\du,19.000000\du);
}
\definecolor{dialinecolor}{rgb}{0.000000, 0.000000, 0.000000}
\pgfsetstrokecolor{dialinecolor}
\node at (27.150000\du,4.272500\du){$\mathcal{H}$};
\pgfsetlinewidth{0.100000\du}
\pgfsetdash{}{0pt}
\pgfsetdash{}{0pt}
\pgfsetmiterjoin
\pgfsetbuttcap
{
\definecolor{dialinecolor}{rgb}{0.000000, 0.000000, 0.000000}
\pgfsetfillcolor{dialinecolor}
\definecolor{dialinecolor}{rgb}{0.000000, 0.000000, 0.000000}
\pgfsetstrokecolor{dialinecolor}
\pgfpathmoveto{\pgfpoint{6.000000\du}{20.000000\du}}
\pgfpathcurveto{\pgfpoint{17.000000\du}{3.000000\du}}{\pgfpoint{19.000000\du}{19.000000\du}}{\pgfpoint{27.000000\du}{14.000000\du}}
\pgfusepath{stroke}
}
\pgfsetlinewidth{0.200000\du}
\pgfsetdash{}{0pt}
\pgfsetdash{}{0pt}
\pgfsetbuttcap
{
\definecolor{dialinecolor}{rgb}{0.000000, 0.000000, 1.000000}
\pgfsetfillcolor{dialinecolor}
\pgfsetarrowsend{stealth}
\definecolor{dialinecolor}{rgb}{0.000000, 0.000000, 1.000000}
\pgfsetstrokecolor{dialinecolor}
\draw (15.000000\du,12.000000\du)--(20.400000\du,6.750000\du);
}
\pgfsetlinewidth{0.200000\du}
\pgfsetdash{}{0pt}
\pgfsetdash{}{0pt}
\pgfsetbuttcap
{
\definecolor{dialinecolor}{rgb}{1.000000, 0.000000, 0.000000}
\pgfsetfillcolor{dialinecolor}
\pgfsetarrowsend{stealth}
\definecolor{dialinecolor}{rgb}{1.000000, 0.000000, 0.000000}
\pgfsetstrokecolor{dialinecolor}
\draw (15.000000\du,12.000000\du)--(21.000000\du,11.250000\du);
}
\pgfsetlinewidth{0.200000\du}
\pgfsetdash{{\pgflinewidth}{0.200000\du}}{0cm}
\pgfsetdash{{\pgflinewidth}{0.200000\du}}{0cm}
\pgfsetbuttcap
{
\definecolor{dialinecolor}{rgb}{0.000000, 1.000000, 0.000000}
\pgfsetfillcolor{dialinecolor}
\definecolor{dialinecolor}{rgb}{0.000000, 1.000000, 0.000000}
\pgfsetstrokecolor{dialinecolor}
\draw (23.000000\du,11.000000\du)--(7.000000\du,13.000000\du);
}
\pgfsetlinewidth{0.100000\du}
\pgfsetdash{{1.000000\du}{1.000000\du}}{0\du}
\pgfsetdash{{1.000000\du}{1.000000\du}}{0\du}
\pgfsetbuttcap
{
\definecolor{dialinecolor}{rgb}{1.000000, 0.000000, 0.000000}
\pgfsetfillcolor{dialinecolor}
\definecolor{dialinecolor}{rgb}{1.000000, 0.000000, 0.000000}
\pgfsetstrokecolor{dialinecolor}
\draw (20.400000\du,6.800000\du)--(20.850000\du,11.350000\du);
}
\definecolor{dialinecolor}{rgb}{0.000000, 0.000000, 0.000000}
\pgfsetstrokecolor{dialinecolor}
\node at (28.550000\du,13.672500\du){$\mathcal{M}$};
\definecolor{dialinecolor}{rgb}{0.000000, 0.000000, 0.000000}
\pgfsetstrokecolor{dialinecolor}
\node at (24.250000\du,10.972500\du){$\mathbb{T}$};
\definecolor{dialinecolor}{rgb}{0.000000, 0.000000, 0.000000}
\pgfsetstrokecolor{dialinecolor}
\node at (15.050000\du,14.322500\du){$\ket{\Psi[\vec{a}(t)]}$};
\definecolor{dialinecolor}{rgb}{0.000000, 0.000000, 0.000000}
\pgfsetstrokecolor{dialinecolor}
\node at (21.200000\du,12.572500\du){$\ket{\Phi}$};
\end{tikzpicture}
  \caption{
    TDVP schematic showing the projection of a tangent vector (blue arrow) in 
    the full Hilbert space $\mathcal{H}$ onto the tangent space $\mathbb{T}$ 
    of a variational sub-manifold $\mathcal{M}$, resulting in a tangent
    vector $\ket{\Phi}$ (red arrow).}
  \label{fig:tdvp}
\end{figure}
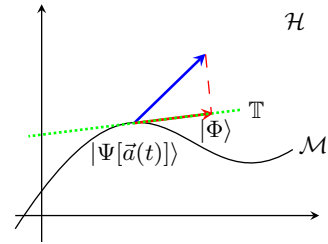

We assume that an exact infinitesimal time step has the form 
$\ket{\Psi(t + dt)} = \ket{\Psi(a(t))} + dt\,O\ket{\Psi(a(t))}$, with some
operator $O \in \mathcal{B}(\mathcal{H})$, which would be the Hamiltonian (times $\mathrm{i}$) 
of the system in the case of unitary dynamics. 
As illustrated in Figure \ref{fig:tdvp}, the locally optimal approximation
to the dynamics is then given by the vector 
$\ket{\Phi(b)} \equiv b^j \partial_{a^j} \ket{\Psi(a)}$,
belonging to the tangent plane $\mathbb{T}$ to the variational manifold $\mathcal{M}$
at the point $a$, which best approximates the vector $O\ket{\Psi(a(t))}$ (blue arrow):
\begin{equation*}
  b = \arg \min_{b'} \left| \ket{\Phi(b')} - O\ket{\Psi(a(t))} \right|^2.
\end{equation*}
Solving this optimization problem is equivalent to numerical integration
of a set of flow equations for $a$
\begin{equation*}
  \partial_t a^j = -g^{jk} \partial_{\overline{a}^k} \bra{\Psi(a)}O\ket{\Psi(a)}, 
\end{equation*}
where $g^{jk}$ is the inverse of the ``metric''
$g_{jk}(\overline{a},a) \equiv \partial_{\overline{a}^j} \partial_{a^k} \langle \Psi(a) | \Psi(a) \rangle$.
In the case where $g_{jk}$ is not invertible, a pseudo-inverse may be used instead.
The TDVP flow equations can also be derived via the principle of least action.
For more details, see for example \cite{haegeman_tdvp_mps} (supplementary material).

\subsection{Derivation of the Fokker-Planck equation for sampling from the Lindblad master equation}

As summarized in the main text, we want to capture the behavior of systems of the form
\begin{align}
 \partial_t \rho &= - \mathrm{i} [K,\rho] - \frac{1}{2}\sum_\alpha({L_\alpha}^\dagger L_\alpha \rho + \rho {L_\alpha}^\dagger L_\alpha - 2 L_\alpha \rho {L_\alpha}^\dagger) \nonumber \\
 &= Q\rho + \rho Q^\dagger + \sum_\alpha L_\alpha \rho {L_\alpha}^\dagger,\label{eqn:supp:lindblad}
\end{align}
with $Q = -\mathrm{i}K - \frac{1}{2}\sum_\alpha {L_\alpha}^\dagger L_\alpha$,
where $K$ is Hermitian and $\rho$ the state that is to be evolved. 
${L_\alpha}$ is the set of arbitrary Lindblad operators that model dissipation.
\begin{equation} \label{eqn:supp:mixed_ansatz}
 \partial_t \rho_t = \int_{\mathcal{M}} \partial_t p_t(\bar{a}, a) \ketbra{\Psi(a)}{\Psi(a)} da d\bar{a},
\end{equation}
where $p_t(\bar{a}, a)$ is a time-dependent probability distribution over the 
pure state variational parameters $a$ and $\mathcal{M}$ is the sub manifold
of Hilbert space formed by the states in the variational class.
Since this integral cannot be (efficiently) performed for a general class
of pure states, we exploit stochastic calculus to sample from it.

First, we review some details of stochastic differential equations (SDE).
Following Gardiner \cite{Gardiner2002}, the expectation value of a 
function of a random variable described by an Ito SDE is
\begin{multline}
 \left\langle\frac{df[x(t)]}{dt}\right\rangle = \frac{d}{dt} \left\langle f[x(t)]\right\rangle \\
 = \left\langle r[x(t),t] \partial_x f + \frac{1}{2}u[x(t),t)]^2 \partial_x^2f\right\rangle,
\end{multline}
where $x(t)$ is a random variable, $r(x, t)$ is the drift coefficient, and $u(x, t)$ is the
diffusion coefficient. If $x$ takes values according to the conditional probability density 
$p(x,t|x_0,t_0)$, for initial conditions $x_0$ at $t_0$, then we can write
\begin{multline}
 \frac{d}{dt}\left\langle f(x[t]) \right\rangle = \int dx f(x)\partial_t p(x,t|x_0,t_0) \\
 = \int dx \left[ r(x,t) \partial_x f + \frac{1}{2} u(x,t)^2 \partial_x^2 f \right] p(x,t|x_0,t_0).
\end{multline}
If we now integrate by parts and discard surface terms, we get
\begin{align} \label{eqn:supp:sde_func_ev}
 \int dx f(x)\partial_tp = \int dx f(x) \bigg\{ &-\partial_x [r(x,t) p] \\
                                        &+ \frac{1}{2} \partial_x^2[u(x,t)^2p]\bigg\}. \nonumber
\end{align}
Choosing $f(x) = 1$, we learn that
\begin{equation}
 \partial_tp = -\partial_x [r(x,t) p] + \frac{1}{2} \partial_x^2[u(x,t)^2p].
\end{equation}
It should thus be clear that the evolution of $p$ is governed by the drift and 
diffusion coefficients $r$ and $u$.

In the case of many complex variables $x^j \in \mathbb{C}$, an It$\overline{\textnormal{o}}$ SDE may take the form
\begin{align} \label{eqn:supp:complex_sde}
  dx^j = r^j(\overline{x}, x, t) dt + U^j_\alpha(\overline{x}, x, t) dw_\alpha,
\end{align}
where $dw_\alpha = \frac{1}{\sqrt{2}}(du_\alpha + idv_\alpha)$ are complex Wiener processes
constructed from real Wiener processes $du_\alpha, dv_\alpha$ such that
$\langle dw_\alpha \overline{dw}_\beta \rangle = \delta_{\alpha \beta} dt$ and $\langle dw_\alpha dw_\beta \rangle = 0$.
One can then derive
\begin{align} 
 \int d\overline{x} dx &f(\overline{x}, x)\partial_tp = \nonumber \\ 
         &\int d\overline{x} dx f(\overline{x}, x) \label{eqn:supp:complex_sde_func_ev}
          \bigg\{ -\partial_{x^j} [r^j(\overline{x}, x,t) p] - \text{c.c.} \\
         & \qquad + \partial_{x^k} \partial_{\overline{x}^l} [U^k_\alpha(\overline{x}, x,t) 
           \overline{U^l_\alpha(\overline{x}, x,t)} p] \bigg\}, \nonumber
\end{align}
where $r^j, U^j_\alpha \in \mathbb{C}$ and $U$ is now called the diffusion matrix.
Again, we can easily obtain the evolution of $p(\overline{x}, x, t)$ by setting $f(\overline{x}, x) = 1$.

We may attempt to find such an equation for $p_t$ in \eqref{eqn:supp:mixed_ansatz},
viewing the entries of $\rho$ as functions of complex random variables $a$ with 
expectation values calculated by integrating over $\mathcal{M}$. In fact, the TDVP
delivers exactly the drift vector and the diffusion matrix needed.
Inserting \eqref{eqn:supp:mixed_ansatz} into \eqref{eqn:supp:lindblad} and using
the TDVP to approximate the vectors $Q\ket{\Psi(a)}$ and $L_\alpha\ket{\Psi(a)}$
as ${b_Q}^j\partial_{a^j}\ket{\Psi(a)}$ and ${b_\alpha}^j\partial_{a^j}\ket{\Psi(a)}$
respectively, we arrive (after partial integration, discarding surface terms) at
\begin{align*}
 \partial_t \rho_t = \int_{\mathcal{M}} \big[ - \partial_{a^j} (p_t {b_Q}^j) - \text{c.c.} 
                     + \partial_{a^k}\partial_{\bar{a}^l}({b_\alpha}^k \overline{{b_\alpha}^l} p_t)\big] \\
                     \times \ketbra{\Psi}{\Psi} da d\bar{a},
\end{align*}
the RHS of which has the same form as \eqref{eqn:supp:complex_sde_func_ev}.
We can thus read off a Fokker-Plank equation for $p_t$
\begin{equation}
 \partial_t p_t = -\partial_{a^j} (p_t {b_Q}^j) - \textnormal{c.c.} + \partial_{a^k} \partial_{\bar{a}^l} \left({b_\alpha}^k \overline{{b_\alpha}^l} p_t\right).
\end{equation}
and obtain an Ito SDE for the variational parameters $a$
\begin{equation}
\label{SDE_new}
  da^j(t) = \left(b_{Q}^j + \langle\bar{L}_\alpha\rangle b_\alpha^j\right) dt + b^j_\alpha dw_\alpha(t).
\end{equation}
Interpreting $a$ as a random, stochastic variable, one could have na\"{i}vely chosen an ansatz of the form $a = b_Q dt + b_\alpha d\omega_\alpha$,
where $d w_\alpha(t)$ is a complex Wiener process (white noise), corresponding
to the linear expression
$d(\ket{\psi}) = Q\ket{\psi}dt + L_\alpha \ket{\psi} d\omega_\alpha$.
This however is unsuccessful, since the differential Ito calculus for $d(\ketbra{\psi}{\psi})$ resolves to
\begin{equation}
 d(\ketbra{\psi}{\psi}) = \ketbra{d\psi}{\psi} + \ketbra{\psi}{d\psi} + L_\alpha \ketbra{\psi}{\psi} \bar{L}_\alpha dt
\end{equation}
and we actually want to evolve the system as a probability density functional instead of a linear Hilbert space vector.
By integrating eq. (\ref{SDE_new}) we can evolve a pure state component 
$\ket{\Psi(a)}$ of some initial $\rho$
such that it samples the evolution of the full mixed state. Using $N$ such samples,
properties of $\rho_t$ can be approximated with an error (variance) that scales as $1/\sqrt{N}$.

\subsection{How to sample from a Wiener process}

It is vital for understanding the presented Monte Carlo scheme that we do not attempt to 
correctly approximate the actual mixed state $\rho$ at some time $t$, but rather the
expectation values of observables of interest. For example, the
observable $\sigma^z$ transforms under the time evolution $\rho(t) = e^{t\LL}(\rho_0)$,
where $\LL$ is the completely-positive trace-preserving map corresponding to 
the RHS of \eqref{eqn:supp:lindblad}, as
\begin{equation}
 \langle\sigma^z\rangle(t) = \tr(\sigma^z e^{t\LL}(\rho_0))
\end{equation}
and the stochastic expectation value for $N$ samples is
\begin{equation}
 \langle\langle\sigma^z\rangle\rangle(t) = \frac{1}{N} \sum_{l=1}^{N} \langle\sigma^z\rangle_l(t),
\end{equation}
where $\langle \sigma_j^z \rangle_l$ is the expectation value of $\sigma_j^z$ for the $l$th pure state sample.

Wiener processes were introduced as a tool to analyze and explain Brownian motion statistically.
For example Kloeden \cite{Kloeden2003} defines a Wiener process $W = {W(t\,|\,t>0)}$ to be a continuous Gaussian process with independent increments such that
\begin{align*}
 W(0) = 0\;\text{a.s.}, &\hspace{.2cm}\langle W(t)\rangle = 0, \\ &\text{and } \textrm{Var}\left(W(t) - W(s)\right) = t-s.
\end{align*}
It can be discretized into $n$ steps with length $\Delta t$ as $W_{n\Delta t} = \sqrt{\Delta t} \sum_{i=1}^n X_i$, 
where $X_i$ are Gaussian random variables.
In this form numerical treatment is easy as long as good Gaussian pseudo-random numbers are available.
In our case we use the NumPy Python framework incorporating the Mersenne Twister MT 19937 algorithm, 
which is widely used for Monte Carlo calculations.\\

\newpage

\subsection{Supplementary plots}

\begin{figure}[h!]
	  \includegraphics[width=.45\textwidth]{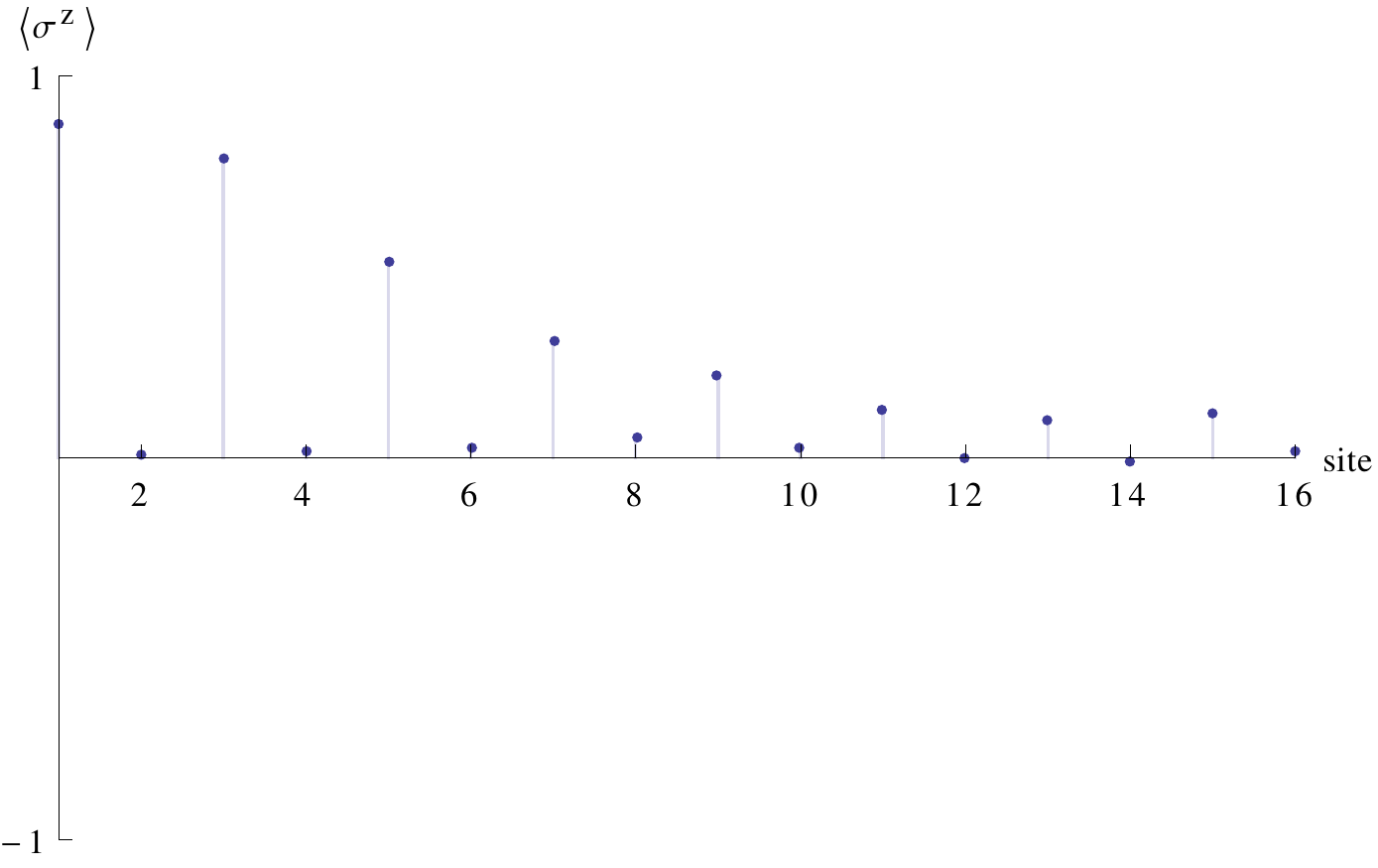}
 \caption{Homogenous dissipation of the form $L_\alpha = \sigma^+_\alpha$ on the Heisenberg KX chain with ($n = 16$, $N = 1000$, $D = 32$).
 The expectation values are skewed because the first spin points up due to initial driving and the last down due to anti-ferromagnetic constraints.
 Note how the absolute values do not become negative because the dissipation is strictly positive.}
  \label{fig:XXZbihom}
\end{figure}

%
%


\end{document}